# Orchestration of Global Software Engineering Projects
## - Position Paper -

Christian Bartelt[1], Manfred Broy[2], Christoph Herrmann[4], Eric Knauss[3],
Marco Kuhrmann[2], Andreas Rausch[1], Bernhard Rumpe[4], Kurt Schneider[3]

[1]*Technische Universität Clausthal, Software Systems Engineering*
*Julius-Albert-Str. 4, 38678 Clausthal-Zellerfeld, Germany*
*{bartelt | rausch}@tu-clausthal.de*

[2]*Techniche Universität München, Institut für Informatik – I4*
*Boltzmannstr. 3, 85748 Garching, Germany*
*{kuhrmann | broy}@in.tum.de*

[3]*Leibniz Universität Hannover, Welfengarten 1, 30167 Hannover, Germany*
*{kurt.schneider | eric.knauss}@inf.uni-hannover.de*

[4]*RWTH Aachen, Lehrstuhl Informatik 3 (Softwaretechnik)*
*Ahornstr. 55, 52074 Aachen, Germany*
*{rumpe | herrmann}@se-rwth.de*

## Abstract

*Global software engineering has become a fact in many companies due to real necessity in practice. In contrast to co-located projects global projects face a number of additional software engineering challenges. Among them quality management has become much more difficult and schedule and budget overruns can be observed more often. Compared to co-located projects global software engineering is even more challenging due to the need for integration of different cultures, different languages, and different time zones – across companies, and across countries. The diversity of development locations on several levels seriously endangers an effective and goal-oriented progress of projects. In this position paper we discuss reasons for global development, sketch settings for distribution and views of orchestration of dislocated companies in a global project that can be seen as a "virtual project environment". We also present a collection of questions, which we consider relevant for global software engineering. The questions motivate further discussion to derive a research agenda in global software engineering.*

## 1. Introduction

Today's IT and Software industries are pretty globally distributed. Globally distributed projects are rapidly becoming the norm for the development of large software systems. It is no longer unusual for a large project to have teams in more than one location, often on more than one continent. Therefore software engineering has to cope with distributed execution and local management but integrated, global project solutions.

### 1.1 Promises of Global Software Engineering

Global software engineering (GloSE) is a consequence of a variety of current trends and profane necessities. These reasons for global software engineering can all be reduced to the following *three main significant forces* that have pushed global software engineering as a fact in our daily work:

- *Economically*: Those reasons include e.g. cost concerns, like *significant differences in personnel costs*, if for example the development in Asia is dramatically cheaper than in Europe. Other samples for economic reasons are the increasingly global networks of companies to develop increasingly complex software under (time and budget) pressure and in competition to each other. Globally distributed development promises chances for being better than the competitors.
- *Organizationally*: Organizational reasons can be motivated by the structure of globally acting companies. If a company is spread over the whole world, distributed development is the *natural style of project organization* as development resources are already located multi-sited. Another typical organizational reason is the need to tap global

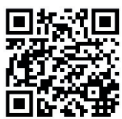

[BBH+09] C. Bartelt, M. Broy, C. Herrmann, E. Knauss, M. Kuhrmann, A. Rausch, B. Rumpe, K. Schneider
Orchestration of Global Software Engineering Projects
In: Proceedings of the Third International Workshop on Tool Support Development and Management in Distributed Software Projects,
collocated with the Fourth IEEE International Conference on Global Software Engineering ICGSE 2009, July 13-16 2009, Limerick, Ireland
www.se-rwth.de/publications

pools to acquire highly skilled *resources* and finding an appropriate mix of expertise for a project.
- *Strategically*: Another face is given by strategic reasons for distributed development. If a globally acting software vendor like SAP, Siemens, or Microsoft produces localized software, the location of developers close to the customers could have advantages: Culture is known, time to market can be optimized, e.g. for localized patches, updates etc. or *psychological/political aspects* are taken into account due to local employees. Satisfying investment requirements imposed by governments in foreign markets, mergers, and acquisitions are also examples for strategic reasons for global software engineering.

## 1.2 Applying GloSE in Practice

Despite the stated reasons there are numerous ways of applying global software engineering in practice. They reach from simple buyer-supplier relationships to sophisticated peer development in specialized areas. Each *distribution setting* has consequences especially related to project organization/coordination and collaboration. A distribution setting can be motivated according the above stated reasons, where a mixture of those reasons is also possible. On the one hand the economic driver can locate "simple" development tasks at Asia while formidable (expensive) tasks like requirements engineering or architectural design stay in Europe. On the other hand design and development can be distributed among partners by assigning sub-tasks. These two settings have completely different implications.

Such differences in distribution settings have strong impacts on a particular project or even on the whole organization. Referring to cost-driven phase-based task distribution, the following setting is usual: Development teams grew up, as sub-projects in the outsourcing countries also need some organization and so on. Beside the pure development, management overhead occurs. Looking at the second setting distributed and collaborative work within one discipline (e.g. architectural design), infrastructure requirements (e.g. for "virtual white boards") may increase. Corresponding to the concrete setting, a variety of organizational measures have to be prepared – with consequences to the project, its organization, coordination etc.

## 1.3 Results of GloSE in Practice

Independent from the co-located software engineering approach, global software engineering is becoming the predominant way of software engineering. It has also numerous additional risks to classical software engineering, like the variability of the project settings, distances between participants and resulting *professional and social issues*.

Because of widely spread projects, spanning several countries and cultures, projects became "multicultural". This challenge is independent of the concrete reason of distribution. A similar challenge is distance in time and space, a major problem, although it may seem minor at first glance. Different time zones hamper synchronous communication and delay coordination in a project. The inability of face-to-face meetings has decreased communication richness [1].

Originally, global development was intended to reduce costs. But the result was that many projects showed the same symptoms related to quality or communication-lacks as one-site projects [7]. In fact the consequences were even worse: Each communication-issue, each under-specified component caused increased effort, leading even to *increased* costs. Especially if the driver was not organizational nor strategic but economic, the (economic) success of the project is jeopardized independent from the quality of the deliveries. Tasks in global software development projects often take much longer than in co-located environments [3] and suffer from a wide range of problems [6].

## 1.4 What is GloSE All About?

Apart from the forces that push global software engineering and from the obvious problems and drawbacks, there is little reason to expect global software engineering to be diminished in the future. Rather, it appears that we face increasing globalization of markets and production, increasing the pressure to distribute projects globally and thus broaden the appearance of global software engineering.

> *Our Notion of Global Software Engineering:*
>
> We want to apply the concepts of software engineering to the benefit of global projects. However, there seem to be specific obstacles. A key question is: What is the difference between classical software engineering and global software engineering?
>
> In global software engineering work is allocated to people at ***distributed sites*** with ***different software engineering cultures***.

This notion of global software engineering also provides the explanation for the problems that appear in applying global software engineering in practice:

In software development projects the tasks cannot be seen as isolated activities. There exist complex dependencies between particular tasks. Thus, people have to *communicate* with each other to fulfill their tasks. If tasks are carried out at distributed sites, people at any given site have to communicate with each other. But in a global environment communication suffers [6].

In a traditional, co-located project, teams usually have naturally built up a number of ways of coordinating their work. They have a shared *view* of how the work will proceed, either because of a shared, defined process or just by acquiring a common set of habits and vocabulary over time. The difference in global software engineering is that many of the mechanisms for coordinating the work in a co-located setting are absent or disrupted. In global software engineering people with different software engineering cultures work together. This generates a higher demand to *co-ordinate* the different tasks with each other. Even worse, in a global environment such coordination is more difficult than in a co-located setting [8].

---

*The Main Challenge in GloSE:*

Thus, the key challenge of global software engineering is to **establish appropriate communication and coordination habits** in a global project environment (see also [9]).

---

### 1.4 Scope and Claim of our Approach

Communication and coordination in global software engineering can be investigated from two different views: From *social* aspects and *professional* aspects. In this paper we focus on the professional aspects of global software engineering.

---

*Our Approach for Successful GloSE*:

To be successful in global software engineering one has to (**re-)orchestrate the existing communication and coordination cultures of all parties participating** in the global software engineering project.

---

This (re-)orchestration has to be established on three levels: *Project set-up and management*, *processes and information flows* and *artifacts and product models*. On each level the organizational, e.g. establishing specific handshake tasks between people working together but having different software engineering cultures in mind, as well as the technical aspects, e.g. providing communication infrastructures for a global project environment, have to be improved. The three levels will be more detailed discussed in the following section.

## 2. A General Approach for Orchestration of Global Software Engineering Projects

As there are manifold reasons to establish distributed development, the main problems on the professional level are an unclear and undefined coupling between the distributed organizations and locations and in consequence missing knowledge and practice for the global project as a whole. One reason therefore is a vague understanding of interfaces (e.g. data dependencies, process connections) between the distributed locations. We consider the lack of explicit interfaces a critical issue; it resides on the levels of integration mentioned above (project set-up, processes, and artifacts). We consider the re-orchestration of these three layers – including organizational as well as technical questions – a fundamental challenge. We propose an approach covering communication, process, and technology. In detail:

- Tracing and consistency controlling of multi-sited dependencies of data and information
- Constitution of a multi-site GloSE process by integration of organization-specific processes
- Constitution of a multi-sited GloSE project organization with respect to organization-specific structures

### 2.1 Example and Discussion

Figure 1 provides a sample of globally acting companies and their cooperation. We consider two companies *Organization A* and *Organization B*, who work together in a *Virtual Project*.

Both have existing project teams, processes and data storage structures. Due to a distributed project, *sub-sets* of personnel and data structures have to be combined as well as the development processes. Those elements become *visible* in the global context. As shown in Figure 1 (left column) the virtual team is built of selected members of both companies. Both have also "internal" supporting staff that is not visible in the virtual context. Also shown is the orchestration of the development processes. So each individual process may contain steps not available in the other processes to be considered. As shown in Figure 1 (middle column) several steps are unique for one site, so processes are orchestrated. On the other hand, particular steps may also be integrated (e.g. A.5 and B.5 to an integrated 5 on the virtual project's level). Third the data storages are connected in the virtual project. Each site has its own storage, processes, and team-structure.

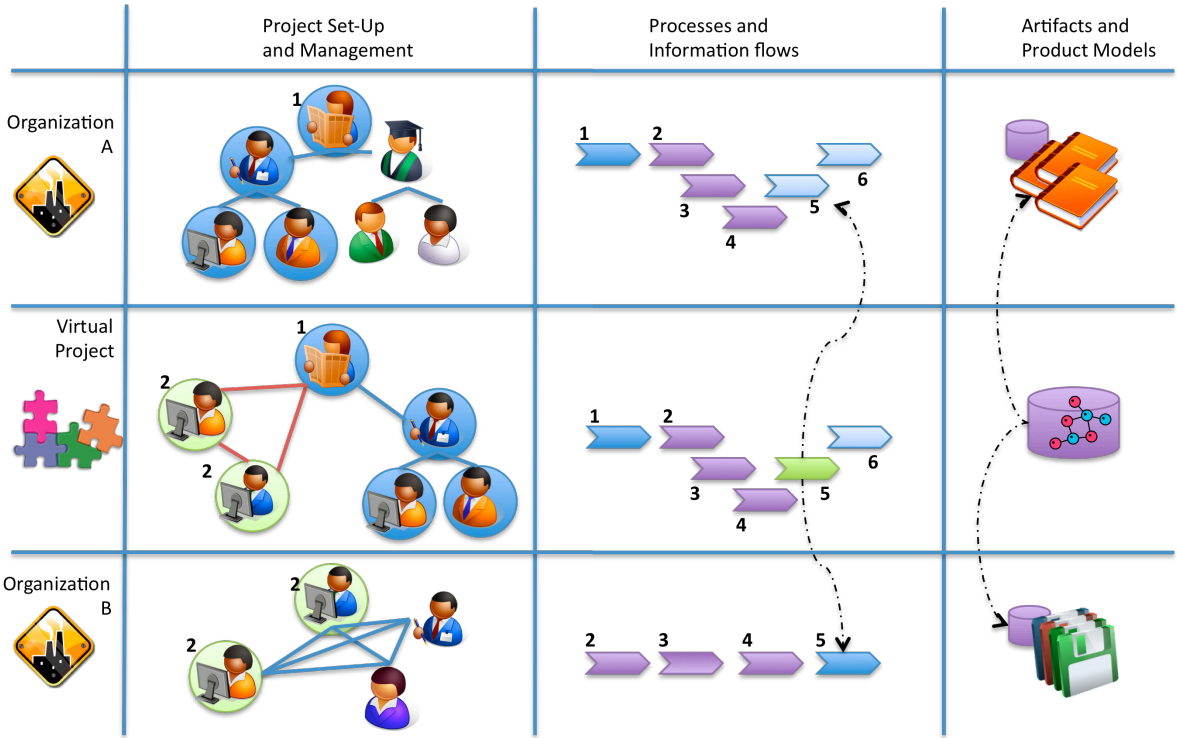

**Figure 1 Example: Virtual Project setting for orchestration and integration in GloSE**

On top the virtual project provides several conceptual / virtual *views* on the site-specific artifact-, process-, and organizational-models. The site-specific views remain, because they allow specialized teams to work in their ideal environment. The combined views are necessary for the project, because a common understanding of the shared concepts is needed. The elements for the shared views are collected from (all) participating sites and integrated to task-specific views, e.g. several codes for a specific task and the specifications the code belongs to for reviewing tasks. Other examples include so-called *Dashboards* for project management [5] that collect and provide information of the distributed project to the managers. For GloSE this is not enough as not only the collection and presentation of data is the matter but also the collaborative creation of and work on shared artifacts.

## 2.2 Views in GloSE

As projects in a global context are considered loosely coupled, the proposed *virtual project* on the one hand realizes (technical) orchestration and integration aspects. On the other hand, views are provided. A view is a *site-independent*, *task-driven* and *role-specific* snapshot of the project.

A view is *site-independent* as selected items are part of the view independently of their "location". Referring to Figure 1 development artifacts can be the matters, which are provided by several sites (e.g. specifications from *Organization A* and corresponding source codes from *Organization B*). They together build a part of the virtual project's artifact structure.

A view is *task-driven* as selected items are collected and presented according to the current task. A developer, who shall implement a particular requirement, will only see code and corresponding specifications (requirements, tests etc.). Non-development artifacts or artifacts not relevant for that very task are out of scope.

A view is *role-specific* as elements are selected respecting the needs of being informed according to a role and its tasks or responsibilities in the project. A project manager for instance would be interested in all status information of the *virtual project*. The required information in that view is presented respecting the current *virtual development process* and all relevant artifacts from all sites.

Thus views are a fundamental concept for virtual projects in global software engineering, definition of views is a challenging task. As a first step, we start to identify, orchestrate and integrate typical subjects relevant for (distributed) development. Relevant subjects are processes and organization structure, information flows, and artifacts (also refer to Figure 1).

### 2.2.1 Considering Processes

A fundamental challenge is the handling of distributed and heterogeneous development processes. Each organization taking part in a distributed project usually has its own accepted development process. Those processes either need to be harmonized (especially accompanying ones) or coupled in some way. It can be even more demanding to identify differences in the first place. This is, however, a precondition to communicate and agree on interfaces on each level. In the above example *A* and *B* have different communication practices [2]. If developers at *A* and *B* are not made aware of those differences, they will not understand and accept a modification to their way of working.

Accompanying processes mainly target organizational matters. So on the one hand process interfaces provide the specifications how to couple organizations. On the other hand interfaces have to be considered together with organizational questions related to the whole (global) project.

> *The challenging task is to identify the integration options, appropriate process-interfaces or to define some kind of common development process. The harmonization of processes is necessary to build a common understanding of the whole global project. Understanding means a common vocabulary (terminology, ontology), a common set of milestones, deliverables, common strategies for coordinating the distributed (sub-)projects and knowledge of the requirements related to process-relevant artifacts*

### 2.2.2 Considering Artifacts

Another challenge in distributed development is the management of development artifacts. Those artifacts are distributed by the nature of the project setting. In fact all artifacts in a project are interdependent in several ways. For example, certain architecture specifications motivate the creation of particular software components. This way a dependency between the components (the code) and the specification exists.

> *The questions concerning artifacts are: Who owns what artifact? Are the artifacts consistent with each other? Are there redundancies e.g. because of the specification is mirrored at the developers' location? And if so, are both copies of the specification consistent?*

An important requirement is the transparency of those artifacts to make sure that the development work matches first the specification and second the project's goals.

### 2.2.3 Considering Project Organization

The third challenge we want to outline is the organization structure of sub-projects in distributed settings. A communication net, where everybody talks to everybody else is inefficient and will not work in distributed settings. Furthermore it is necessary to determine existing organization structures and to identify a suitable integration structure of sub- or sub-sub-projects into the whole distributed setting.

> Questions to be answered are: *What are the responsibilities in the particular projects? What are the communication paths? Is there a correlation between communication and responsibility? Is there some kind of "virtual super project"?*

### 2.2.4 Consequences

The questions sketched above shortly outline the main problems that can be found when combining organizations within a distributed development project. In fact those issues may result in serious problems, beginning at misunderstandings related to requirements up to permanent communication lacks if different cultures are disobeyed. Those gaps lead to quality and efficiency issues with *increased costs in consequence*. The fragmentation of artifacts, processes and project organization in distributed projects are *project risks*. We also started to discuss questions related to issues that go beyond "pure" integration. The scope was not a set of particular sub-projects but the whole distributed ("virtual") project. Questions refer namely to coordination and responsibilities. Nevertheless orchestration and integration on different levels are in our opinion an adequate way to realize risk-oriented project management for selected settings.

## 3. Conclusion and Further Issues

In this discussion paper we have identified the reasons for the main challenges of global software engineering. We have identified typical problems on three levels and an initial set of core questions for outlining problems and matters to be considered. This provides a baseline for discussing typical problems of global software engineering. In the following we state some additional questions that need further research:

- Are there further relevant critical problems in GloSE projects, besides the identified integration problems?

- Does our form of intended orchestration of projects cover all relevant aspects and which additional problems will be encountered?
- Do the three mentioned levels (see sec. 2) cover the whole area of orchestrations/integration capabilities in global software engineering?
- What is the best resp. most efficient way for the realization of the GloSE infrastructure (related to 3.1) in practice?
- With respect to the three levels: Can we address integration issues of each of the three levels separately, or do we need to integrate on all three areas in parallel?
- Are there further and *resilient* experiences in "integrating" technical and non-technical issues, such as communication habits? And what are resulting views?
- What are appropriate techniques for identifying individual practices on each level – without interrupting development work?
- How can the inherit dynamics of a multi-site project be accommodated by our approach – what are the modes of change for an interface?

Those questions aim at the problem of organizing and coordinating a distributed project. We assume that integration *not only solves problems, but also will possibly create new ones*. Furthermore not all aspects of all sub-projects are suitable or even necessary for integration. So the optimal amount of elements has to be determined. Beyond integration, the definition of *views* is necessary – so: what are the views relevant for each project? Can we find common and therefore standardizable patterns of views?

Also to be considered in this context is the question, if there exist processes or artifacts in the global project that span all (or almost all) sub-projects. If so, what are adequate instruments to extract those elements from single projects and handle them on the global level? The possible range for solutions is from (lazy) simple mappings to (strict) contract-based integration. The same question is valid for the tool viewpoint: If connected organizations each have individual tools, what does a GloSE infrastructure looks like? Simple tool-supported features like task management have to be revised in that context [4]. Another question is for the optimal way of enabling organizations for being distributed players. Is it enough simply to change the way of storing artifacts or introduce new processes? As we know from process improvement, investments in only one dimension usually do not have the expected impacts. So is it the same for making organizations GloSE-ready? Or is a weighted strategy incorporating several aspects and dimensions of advantage?

## 5. References

[1] Cockburn, A.: *Agile Software Development*. Addison Wesley, 2002.
[2] Schneider, K., K. Stapel, et al.: *Beyond Documents: Visualizing Informal Communication*. Third International Workshop on Requirements Engineering Visualization (REV 08), Barcelona, Spain, 2008.
[3] Herbsleb, J.D. and Mockus, A.: *An Empirical Study of Speed and Communication in Globally-Distributed Software Development*. IEEE Transactions on Software Engineering, 29, 3 (2003), p. 1-14.
[4] Kuhrmann, M., Kalus, G. and Chroust. G.: *Tool-Support for Software Development Processes*. In Enterprise Information Systems for Business Integration in SMEs: Technological, Organizational and Social Dimensions, to appear 2009.
[5] Münch, J. and Heidrich, J.: *Software project control centers: concepts and approaches*. Journal of Systems and Software 70(1-2), p. 3-19, (2004.
[6] Olson, G.M. and Olson, J.S., *Distance Matters*. Human-Computer Interaction, 15, (2000), p. 139-178.
[7] The Standish Group: *Chaos Report*. 2006.
[8] Whitehead, J.: *Collaboration in Software Engineering:A Roadmap*, in *Future of Software Engineering 2007*, L.Briand and A. Wolf, Editors. 2007.
[9] Herbsleb, J. D.: *Global Software Engineering: The Future of Socio-technical Coordination*. 2007.